\documentclass[aps,prl,preprint,superscriptaddress]{revtex4}

\begin{document}


\title{An example of  Kaluza-Klein-like theories leading after compactification to massless spinors coupled
to a gauge field}
\author{N. Manko\v c Bor\v stnik}
\affiliation{Department of Physics, University of
Ljubljana, Jadranska 19, SI-1111 Ljubljana}
\altaffiliation{Primorska Institute for Natural Sciences and Technology,
C. Mare\v zganskega upora 2, SI-6000 Koper}
\author{H. B. Nielsen}
\affiliation{Department of Physics, Niels Bohr Institute,
Blegdamsvej 17, Copenhagen, DK-2100}


\author{}
\affiliation{}


\date{\today}

\begin{abstract}
The genuine Kaluza-Klein-like theories (with no fields in addition to gravity with torsion) have difficulties 
with the existence of massless spinors after the compactification
of some of dimensions of space\cite{witten}. We demonstrate in this letter on an example of a flat torus - as 
a compactified part
of an (1+5)-dimensional space - that for constant fields and appropriate  boundary conditions there exists 
in the $1+3$ dimensional space a massless 
solution, which is mass protected and    chirally coupled  
with a Kaluza-Klein charge to the corresponding gauge field. 

\end{abstract}


\maketitle

{\it Introduction:}
{\it Genuine Kaluza-Klein-like theories}, assuming nothing but  a gravitational field 
in $d$-dimensional space (no additional gauge or scalar fields), 
which after the spontaneous compactification of a $(d-4)$-dimensional part of space manifest in 
four dimensions as all the known gauge fields including gravity, have difficulties\cite{witten} with 
masslessness of fermionic fields at low energies. It looks  
namely very difficult to avoid after the compactification of a part of space  the appearance of  
representations of both handedness in this part of space and consequently also in the 
(1+3)-dimensional space. Accordingly, the gauge fields can hardly couple chirally in the (1+3) - dimensional 
space. 

In more popular versions, in which one only uses the idea of extra dimensions but does not use
gravity fields themselves to make gauge fields, by just having gauge fields from outset, the break of the parity 
symmetry in the compactified part of space is achieved, for instance, by (an outset of) magnetic fields\cite{salamsezgin}. 
Since gravity does not violate parity, also typically not in extra dimensions alone, it looks accordingly 
impossible to make the genuine Kaluza-Klein gauge particles coupled chirally\cite{holgercolinbook,witten}.
The most popular string theories, on the other side, have such an abundance of ``fundamental`` (or rather 
separate string states) gauge fields, that there is (absolutely) no need for the genuine Kaluza-Klein ones.

In an approach by one of us\cite{pikanorma,norma} (which is offering also the mechanism for generating families)
it has long been the wish to obtain the gauge fields from only gravity, so that ''everything'' would become gravity.
This approach has taken the inspiration from looking for unifying all the internal degrees of freedom, that is 
the spin and all the charges into only spins. This approach is also a kind of the genuine Kaluza-Klein theory, suffering
the same problems, with the problem of getting chiral fermions included, unless we can solve them.

It is the purpose of this letter to demonstrate on an example of a torus-shaped compactifying space (a flat $M^{1+5}$
space with the $SO(1,5)$ symmetry compactifying into a flat $M^{1+3}$ part of space with the $SO(1,3)$
and $S^1 \times S^1$ symmetry)
that using a gravitational field with a vielbein and a spin connection  and a torsion on a torus 
and choosing appropriate  boundary conditions on a torus, we
indeed get a  genuine Kaluza-Klein originating gauge fields chirally coupled to a masless and mass protected spinor! 

We assume a  real action for a free gravitational field, which is linear in the Riemann scalar and in first 
order formulation. In a two dimensional 
compactified part of space, the Euler-Lagrange equations of motion for a free 
gravitational field with zweibein and spin connection fields give no conditions on  the fields. 
It is the requirement 
that a covariant derivative of vielbeins is zero, which relates  spin connections,  vielbeins and a torsion.

We allow the spin connection fields and consequently also the torsion fields to be complex. 
The action for a spinor field, which interacts with a background gravitational field, 
is  assumed to be  real.
Consequently, the Lagrange density for a spinor field is real and the corresponding Weyl equations 
of motion are Hermitean. It then follows that a spinor field ''sees'' 
either only a real part
or only an imaginary part of components of the spin connection fields, depending on the index of fields. 
On a two dimensional flat torus, the spinor ''sees''  zweibein fields as 
imaginary fields, while  {\it the spinor Lagrange density is not invariant under the gauge 
transformations, generated by the spin generators}.

We demonstrate that for any choice of fields the massless solutions of both handedness exist with the
same Kaluza-Klein charge. To assure mass protection mechanism,  
we make a choice of {\it boundary conditions, which distinguish between the two handedness}. 
The corresponding Weyl equations can then have  massless 
solutions of only one handedness, they are charged with a Kaluza-Klein charge
and also chirally coupled to the corresponding Kaluza-Klein $U(1)$ gauge field, 
when such a gauge field is switched on in $(1+3)-$dimensional ''physical'' space (as a background field). 
 The  Kaluza-Klein charge, which is proportional to the expectation values of the momenta of 
  massless spinors in the two extra dimensions, are expressible with the spin connection fields and the
 corresponding operators.

{\it Weyl spinors in gravitational fields with spin connections and vielbeins:} 
We let a spinor interact with a gravitational field through vielbeins $f^{\alpha}{}_{a}$ (inverted vielbeins to 
$e^{a}{}_{\alpha}$ with the properties $e^a{}_{\alpha} f^{\alpha}{}_b = \delta^a{}_b,\; 
e^a{}_{\alpha} f^{\beta}{}_a = \delta^{\beta}_{\alpha} $ ) and   
spin connection fields\footnote{We assume the signature $\eta^{ab} =
diag\{1,-1,-1,\cdots,-1\}$.}, namely 
$\omega_{ab\alpha}$, which is the gauge field of $S^{ab}= \frac{i}{4}(\gamma^a \gamma^b - \gamma^b \gamma^a)$.
We choose the basic states in the  space of spin degrees of freedom to be 
eigen states of the Cartan sub algebra of the operators: $S^{03}, S^{12}, S^{56}$.

The covariant momentum of a spinor is
\begin{eqnarray}
p_{0 a} = p_{a} - \frac{1}{2} S^{cd} \omega_{cda}, \quad p_{0 a} = f^{\alpha}{}_{a} p_{0 \alpha}.
\label{covp}
\end{eqnarray}
The corresponding Hermitean Lagrange density ${\cal L}$  has for  real vielbeins $f^{\alpha}{}_{a}$ a form
${\cal L} =  \frac{1}{2} [(E\psi^{\dagger}\gamma^0 \gamma^a p_{0a} \psi) + (E\psi^{\dagger} \gamma^0\gamma^a p_{0 a}
\psi)^{\dagger}]$
\begin{eqnarray}
{\cal L} &=& \psi^{\dagger}\gamma^0 \gamma^a  [ E ( p_{a} - \frac{1}{2} S^{cd}  \Omega_{cda})
+ \frac{i}{2} (E f^{ \alpha}{}_{a})_{, \alpha}]\psi,
\label{weylL}
\end{eqnarray}
with $ E = \det(e^a{}_{\alpha}) $, 
$ \Omega_{cda} =\frac{1}{2}( \omega_{cda} + (-)^{cda}  \omega^*{}_{cda})$, and with
$(-)^{cda}$, which is $-1$, if two indices are equal, and is $1$ otherwise (if all three indices are different). 
In our case of $1+5$ space, when two dimensions are compactified into $S^1 \times S^1$ flat torus, and when the
Lagrangean of only the two compactified dimensions is concerned, $(-)^{cda} =-1.$ 
The expressions for spin connection fields $\Omega_{abc}$ demonstrate, 
that fields have either real or imaginary components. In $d=2$ case
$\Omega_{abc}$ is always pure imaginary. One should notice that the Lagrangean (\ref{weylL}) {\it is not invariant
under the gauge transformations of the type } $\psi' = e^{i \alpha_{cd} S^{cd}} \psi$, for a general 
$\alpha_{cd}$ 
(for the same reason some of $\Omega_{abc}$'s are imaginary).

Latin indices  $a,b,..,m,n,..,s,t,..$ denote a tangent space (a flat index),
while Greek indices $\alpha, \beta,..,\mu, \nu,.. \sigma,\tau ..$ denote an Einstein 
index (a curved index). Letters  from the beginning of both the alphabets
indicate a general index ($a,b,c,..$   and $\alpha, \beta, \gamma,.. $ ), 
 from the middle of both the alphabets   
 the observed dimensions $0,1,2,3$ ($m,n,..$ and $\mu,\nu,..$), indices from the bottom of the alphabets
 indicate the compactified dimensions ($s,t,..$ and $\sigma,\tau,..$).

The Lagrange density (\ref{weylL}) leads to the Hermitean Weyl equation
\begin{eqnarray}
\gamma^0 \gamma^a  P_{0a}\psi =0, \quad P_{0a}=  E (f^{\alpha}{}_{a} p_{\alpha} -\frac{1}{2} S^{cd} \Omega_{cda}) 
+ i (E f^{\alpha}{}_{a})_{,\alpha}.
\label{Weylgen}
\end{eqnarray}
Taking now into account that $\gamma^a \gamma^b = \eta^{ab}- 2i S^{ab}$,  $\{\gamma^a, S^{bc}\}_- = i
(\eta^{ab} \gamma^c - \eta^{ac} \gamma^b)$,  one easily finds that\footnote{
 $[a  ...b]$ means, that the expression must be anti symmetrized with respect to $a,b$. }
 $ \gamma^a P_{0a} \gamma^b P_{0b} = Ef^{\alpha}{}_{a}p_{0\alpha} Ef^{\beta a} p_{0\beta} 
+ \frac{1}{2} S^{ab}E^2  S^{cd} {\cal R}_{abcd} + 
S^{ab} {\cal T}^{\alpha}{}_{ab} P_{0 \alpha} + i S^{ab} \frac{1}{2} E f^{\alpha}{}_{[a}(E 
f^{\gamma}{}_{b]})_{,\gamma ,\alpha}.$ We find  
${\cal R}_{ab cd} = f^{\alpha}{}_{[a} f^{\beta}{}_{b]} (\Omega_{cd \beta,\alpha } + \Omega_{ce\alpha}
\Omega^{e}{}_{d \beta})$ and for the torsion:
${\cal T}^{\alpha}{}_{ab} =  
E f^{\alpha}{}_{[a}\{(E f^{\beta}{}_{b]})_{,\alpha} + \Omega_{cb]\alpha}Ef^{bc}\}.$

The most general vielbein for $d=2$ can be written by an appropriate parameterization as
\begin{eqnarray}
e^s{}_{\sigma} = e^{\varphi/2}
\pmatrix{\cos \phi   & \sin \phi \cr
- \sin \phi & \cos \phi \cr}, 
f^{\sigma}{}_{s} = e^{-\varphi/2} \;
\pmatrix{\cos \phi   & -\sin \phi \cr
\sin \phi & \cos \phi \cr},
\label{f1and1}
\end{eqnarray}
with $s=5,6$ and $\sigma =(5),(6)$ and $g_{\sigma \tau} = e^{\varphi} \eta_{\sigma \tau}, g^{\sigma \tau} = e^{- \varphi}
\eta^{\sigma \tau}$, $\eta_{\sigma \tau} = diag(-1,-1) = \eta^{\sigma \tau}$.
If there is no dilatation (which means that the torus is flat) then $ E =1$.

If in the case of $d=2$, the Einstein action for a free gravitational field 
$S= \int \; d^d{} x \; E \; R ,$ with $  
R = f^{\sigma [s} f^{\tau t]} \;\Omega_{st \sigma,\tau} 
$ is varied 
with respect to both, spin connections and vielbeins, the corresponding equations of motion
bring no conditions on any of these two types of fields,
so that any zweibein and any spin connection can be assumed. The only limitation
on zweibeins and spin connections might come from boundary conditions. 
If we require that the total covariant derivative of a vielbein $e^s{}_{\sigma}$ is equal to
zero 
\begin{eqnarray}
e^s{}_{\tau ; \sigma} = 0 = e^s{}_{\tau , \sigma} + \Omega^a{}_{t \sigma} e^t{}_{\tau} - 
\Gamma^{\tau'}{}_{\tau \sigma} e^s{}_{\tau'},
\label{covderviel}
\end{eqnarray}
the spin connections $\Omega^s{}_{t \sigma}$ and 
vielbeins determine the quantity
$\Gamma^{s}{}_{\sigma \tau} $, whose anti symmetric part is a torsion 
${\cal T}^{s}{}_{\sigma \tau} = \Gamma^{s}{}_{\sigma \tau} - \Gamma^{s}{}_{\tau \sigma},$
in agreement with the definition of  the torsion above (${\cal T}^{s}{}_{\sigma \tau} =   
e^{s}{}_{[\sigma,\tau]} +\Omega^s{}_{t[\sigma}e^{t}{}_{\tau]}$). 
Accordingly,  Eq.(\ref{covderviel})
{\it can be for $d=2$  understood as the defining equation 
for ${\cal T}^{\sigma}_{\tau \tau'}$,}
for any spin connection $\Omega_{stu}$ and any zweibein. 
 Because of these facts, the Euler index
\begin{eqnarray}
{\rm Euler}\;{\rm index} &=& \int \; d^d{} x \; E \; {\cal R}, 
\label{Eulerindex}
\end{eqnarray}
 with 
${\cal R} = {\cal R}^{\alpha}{}_{\beta \delta \gamma} \delta^{\delta}{}_{\alpha} g^{\beta \gamma} = 
(\{{}^{\alpha}_{\beta \gamma}\}_{, \delta}
- \{{}^{\alpha}_{\beta\delta}\}_{,\gamma} ) \delta^{\delta}{}_{\alpha} g^{\beta \gamma}, 
$
{\it is not in general equal} to the Euler index computed from  $\Omega_{st t'}$ .
These  two ${\cal R}$ and $R$ are independent\cite{Deser}.

Taking into account Eqs.(\ref{f1and1},\ref{covderviel}) 
we find for the torsion ${\cal T}_{\sigma}: = {\cal T}^{\sigma'}{}_{\tau \tau'}
\varepsilon^{\tau \tau'}
g_{\sigma' \sigma}$ the expression\footnote{We are kindly reminding the reader that the detailed derivations 
of the above expressions, as well of
all further expressions, can be found in refs.\cite{HNP04}.}
\begin{eqnarray}
{\cal T}_{\sigma} = \phi_{; \sigma} + \varepsilon_{\sigma}{}^{\tau} \frac{1}{2} \varphi_{,\tau}, \quad
{\rm with} \; \phi_{; \sigma} = \phi_{, \sigma} - \Omega^{5}{}_{6 \sigma}, \quad {\rm and} \; 
{\cal T}_{[\sigma, \tau]} = - \Omega ^{5}{}_{6 [\sigma, \tau] }. 
\label{Tsigma}
\end{eqnarray}
$\varepsilon_{\sigma}{}^{\tau} $ is the antisymmetric tensor.

{\it Weyl spinors in  $M^2$:}
The Weyl spinor  wave functions $\psi$, manifesting masslessness
 in the $(1+3)$ space, 
must in $M^2$  obey the Weyl equations of motion (Eq.(\ref{Weylgen})). 
 
 The equations of motion for the  massless spinors of the right ($\psi_+$) and the left ($\psi_-$) handedness
 ($\Gamma^{(2)} = -2 S^{56}$),  
 $ S^{56}\psi_{\pm} = \pm \frac{1}{2} \psi_{\pm}$, put into the background field with no dilatation are as follows
 \begin{eqnarray}
   \gamma^0 \gamma^5 \{ e^{2i \phi S^{56} }[(p_{0(5)} -2 S^{56} \phi_{,(5)}) + 
   2i S^{56} (p_{0 (6)}-2i S^{56} \phi_{,(6)})]\}\psi_{\pm} =0, \quad
p_{0(\sigma)}= p_{\sigma} - S^{56} i \omega_{\sigma}.
 \label{weyld=2}
 \end{eqnarray}
 We write $\Omega_{5 6 \sigma} = i \omega_{\sigma}$, with $\omega_{\sigma}^* = \omega_{\sigma}$. 
 Requiring that vielbein fields fulfill the isometry relations leads to $\phi_{,\sigma}=0$ 
 ($\phi_{,\sigma}=0$ follows from $\delta f^s{}_{\sigma} = 0 $, 
 see discussions bellow). We then have
 \begin{eqnarray}
   \{[p_{(5)} + \frac{1}{2} \omega_{(6)}] + i[p_{(6)} -\frac{1}{2} 
   \omega_{(5)}]\} \psi_+ =0, \nonumber\\
\{[p_{(5)} +\frac{1}{2} \omega_{(6)}] - i[p_{(6)} -\frac{1}{2} 
   \omega_{(5)}]\} \psi_- =0.
 \label{weyld=2con}
 \end{eqnarray}
 Eqs.(\ref{weyld=2con}) {\it demonstrate} that {\it what ever  $\omega_{\sigma}$  could be,
 there always exist solutions of the Weyl equations of both handedness, which have the same
 $p_{\sigma} \psi$ } ($p_{\sigma} \psi_+ = p_{\sigma}\psi_-$). 
 Since $p_{\sigma}$ determine the Kaluza-Klein charges
 of  spinors ( see Eq.(\ref{kkchargemassless})), it means that massless spinors of both handedness carry the same charge
 and that accordingly the {\it masslessness of spinors is not protected!} 
 
 For the choice: $ \omega_{(5)} = \frac{1}{\pi r} n^{0}{}_{(6)}$, $\omega_{(6)} 
 = -\frac{1}{\pi r} n^{0}{}_{(5)}$, which requires that components of the field are 
 proportional to integers, the solutions of the Weyl equations (\ref{weyld=2con}) have the form
 \begin{eqnarray}
 \psi_{\pm} = e^{-i n^{0}{}_{\sigma} \hat{x}^{\sigma}} \hat{\psi}_{\pm}, 
 \label{solweyl}
 \end{eqnarray}
where $\hat{\psi}_{\pm}$ describe all the other degrees of freedom except the dependence on $x^{\sigma}$ and
$\hat{x}^{\sigma} 2 \pi r = x^{\sigma}$, while $r$ stays for the two radii of the torus. The solutions 
obey periodic boundary conditions.

To assure mass protection we must introduce boundary conditions
which  ''distinguish'' between  spinors of the two handedness.

{\it Boundaries, which allow spinors of only one handedness:} 
We  introduce the action  with boundaries:
\begin{eqnarray}
S &=& \int d^6 x {\cal L} + \nonumber\\  && \int d^6 x \lambda \psi^{\dagger} n_{(5)s} 
n_{(6)t}(\varepsilon^{st}-i\gamma^s \gamma^t) \delta(x^{(5)}-x^{(5)}_{0}) \delta(x^{(6)}-x^{(6)}_{0} )\psi 
\label{actionwithb1}
\end{eqnarray}
with  $n_{(5)s}=(1,0), n_{(6)s} =(0,1)$ and ${\cal L}$ from Eq.(\ref{weylL}). One immediately sees that 
requiring the continuity and differentiability of $\psi$,   
the action (\ref{actionwithb1})  
leads to the solution, for which either 
$\psi$ must on the very point $(x^{(5)}_{0}, x^{(6)}_{0} )$  and accordingly all over the torus be zero 
or $\psi$ must be right handed.  Accordingly, these  
boundary conditions allow only right handed solutions in $d=2$ ($<\Gamma^{(2)}> =<2S^{56}> =1, <\Gamma^{(1+3)}> =
<4i S^{03} S^{12}> = -1, <\Gamma^{(56)}> = <\Gamma^{(1+5)}> =-1$).
Then it follows for the right handed massless solution of Eq.(\ref{solweyl}), for our special choice of 
$\omega_{\sigma} = 
\frac{1}{\pi r} \varepsilon_{\sigma}{}^{\tau} n^{0}{}_{\tau}$, with $\varepsilon_{(5)}{}_{(6)} =1$ 
and $n^{0}{}_{\sigma}$ the two integers that
\begin{eqnarray}
p_{\sigma} \psi_+ = - \varepsilon_{\sigma \tau}
 S^{56} \omega^{\tau} \psi_+.
\label{weylsolr2}
\end{eqnarray}
The right handed solution does not ''feel'' the boundary term at all. 
(We could take as a boundary a wall instead of  one point along one coordinate, 
say along  $x^{(5)}$ at $x^{(6)} = x^{(6)}_0$ and
$x^{(6)}_{0} + 2\pi r.$ Then translational symmetry in one direction ($x^{(6)}$'th) would be broken. For
the spin connection field $\omega_{(5)}=0, \omega_{(6)} = - \frac{1}{\pi r} n^{0}{}_{(5)}$ Eq.(\ref{weylsolr2})
would still be right. The boundary term would now read $\int d^6 x \lambda \psi^{\dagger} n_{(5)s} 
n_{(6)t}(\varepsilon^{st}-i\gamma^s \gamma^t) \delta(x^{(6)}-x^{(6)}_{0} )\psi  +
\int d^6 x \lambda' \psi^{\dagger} n_{(5)s} 
n_{(6)t}(\varepsilon^{st}-i\gamma^s \gamma^t) \delta(x^{(6)}-(x^{(6)}_{0} +2\pi r) )\psi. $
Again spinors of only one handedness would live on such a''torus''.)

We assume that $\omega_{\sigma}$ ''see'' no boundaries.

{\it Spinors coupled to gauge fields in $M^{(1+3)} \times (S^1 \times S^1)$:} 
To study how do spinors couple to the Kaluza-Klein gauge fields in the case of $M^{(1+5)}$, compactified
to $M^{(1+3)} \times (S^1 \times S^1)$, we first look for the appearance of 
pure gauge fields,  when coordinate transformations  of the type $x^{' \mu}= x^{\mu}, x^{' \sigma}= x^{\sigma}
+ \zeta^{\sigma} \theta (x^{\mu})$ are performed and $\delta f^s{}_{\sigma} = 0$ and
$\delta \omega_{\sigma}=0$ is required. 
Under general coordinate transformations $f^{\alpha}{}_{a}$ transform as  vectors with an upper index 
($\delta f^{\alpha}{}_{a} = 
f^{\alpha}{}_{a,\beta} \delta x^{\beta}  + f^{\beta}{}_{a}(\delta x^{\alpha})_{,\beta}$), while
 $\omega^{ab}{}_{\alpha} $  transform as vectors with a lower index
($\delta \omega^{ab}{}_{\alpha} = \omega^{ab}_{\alpha ,\beta}\delta x^{\beta} - 
\omega^{ab}{}_{\beta}(\delta x^{\beta})_{,\alpha}$).
It then follows that {\it $\zeta^{\sigma}$ is a constant,  
$\omega_{\sigma} $ and also $f^{\sigma}{}_{s}$}  
in (Eq.(\ref{f1and1}))
{\it  must be constants}.
 This is in accordance with what we have assumed above for
the spin connection fields - namely that they are constant and even proportional to integers - 
and for vielbeins $f^{\sigma}{}_{s}$ - 
whose derivative we put in Eq.(\ref{weyld=2con}) equal to zero. 
(These assumptions lead to periodic massless spinor solutions on a torus.)
Only constant spin 
connection fields and vielbein fields in $d=2$ fullfil the isometry requirements. 
We start with  vielbeins  $e^m{}_{\mu}=\delta^m{}_{\mu}$ ($f^{\mu}{}_{m}
=\delta^{\mu}{}_{m}$) and 
 $e^s{}_{\sigma} (f^{\sigma}{}_{s})$ from Eq.(\ref{f1and1}) with no dilatation. 

Accordingly we find  $\delta f^{\sigma}{}_{m} = 
\delta^{\mu}{}_{m} \zeta^{\sigma} \theta_{,\mu}, \delta f^{\mu}{}_{s}=0, \delta f^{\sigma}{}_{s} = 0. $ 
Replacing pure gauge fields with
true fields to which these pure 
gauge transformations should belong, we find for  a new vielbein  
\begin{eqnarray}
e^a{}_{\alpha} = 
\pmatrix{\delta^{m}{}_{\mu}  & e^{m}{}_{\sigma} \cr
0= e^{s}{}_{\mu} & e^s{}_{\sigma} \cr},
f^{\alpha}{}_{a} =
\pmatrix{\delta^{\mu}{}_{m}  & f^{\sigma}{}_{m} \cr
0= f^{\mu}{}_{s} & f^{\sigma}{}_{s} \cr}, \quad {\rm with}\; f^{\sigma}{}_{m} = \zeta^{\sigma} A_{\mu} 
\delta^{\mu}{}_{m}
\label{f6}
\end{eqnarray}
and  new spin connection fields 
\begin{eqnarray}
\omega_{56 \mu} =- \zeta^{\sigma} A_{\mu} i \omega_{\sigma}\quad {\rm and} \quad i\omega_{\sigma}.
\label{omega6}
\end{eqnarray}
Here $A_{\mu}$ is the $U(1)$ gauge field, which depends only on $x^{\mu}$.
All the other components of  spin connection fields are zero, since for simplicity we allow no gravity in
$(1+3)$ dimensional space.
The corresponding ${\cal T}^{\alpha}{}_{\beta \gamma}$ transforms as a 
third rank tensor.

To determine the current, coupled to the Kaluza-Klein gauge fields $A_{\mu}$, we
analyze the spinor action
\begin{eqnarray}
{\cal S} = \int \; d^dx E \bar{\psi} \gamma^a P_{0a} \psi = \int \; 
d^dx E \bar{\psi} \gamma^m \delta^{\mu}{}_{m} p_{\mu} \psi + \int \; d^dx E \bar{\psi} \gamma^s f^{\sigma}{}_{s} 
P_{0\sigma} \psi+
\nonumber\\ 
+ \int \; d^dx  E \bar{\psi} \gamma^m \{ f^{\sigma}{}_{m} (p_{\sigma} - S^{56} i \omega_{\sigma} ) -  
\delta ^{\mu}{}_{m} S^{56}  \omega_{56\mu}\}\psi.
\label{spinoractioncurrent}
\end{eqnarray}
$\psi$ defined in $d=(1+ 5)$ dimensional space has the spin part defined in the whole internal space. 
$E$ is for $f^{\alpha}{}_{\beta}$ from Eq.(\ref{f6}) equal to 1.
The first term on the right hand side is the kinetic term ( together with the fourth term represents 
the contribution of the  covariant derivative $p_{0 \mu}$), 
while the second
contributes zero for  massless solutions. Taking into account Eq.(\ref{f6}) and the expression for
$\omega_{56 \mu}$ we end up with only 
the term $\int \; d^dx \bar{\psi} \gamma^m  f^{\sigma}{}_{m} p_{\sigma} \psi$ as a current term.

Since $A_{\mu}$  does not depend on $x^{\sigma}$,  the current in $(1+3)$-dimensional space is for massless spinors
then defined by
\begin{eqnarray}
 A^{\mu} \delta^{m}_{\mu} \int dx^{(5)} dx^{(6)}  \psi^{\dagger} 
 \gamma^0 \gamma^{m} \zeta^{\sigma} p_{\sigma}\psi = - A^{\mu} \delta^{m}_{\mu}\int dx^{(5)} dx^{(6)} \psi^{\dagger}
 \gamma^0 \gamma^{m} \zeta^{\sigma}\Gamma^{(2)} S^{56}\varepsilon_{\sigma}{}^{\tau} \omega_{\tau} \psi , 
\label{KKsolcharge}
\end{eqnarray}

For massless spinors studied above, which solve the Weyl equations of motion (with the boundary included, since
massless spinors of one of the handedness do not even ''feel'' the boundaries), the Kaluza-Klein charges  are equal to
\begin{eqnarray}
Q^{\pm} = \zeta^{\sigma} <\pm |p_{\sigma} |\pm> = \zeta^{\sigma} < \pm |- \Gamma^{(2)} S^{56} 
\varepsilon_{\sigma}{}^{\tau} \omega_{ \tau}
|\pm>,
\label{kkchargemassless}
\end{eqnarray}
where $<\pm||\pm> $ means that the integration over the two coordinates  and
the  corresponding spin degree of freedom is performed for spinors of the  two handedness $\psi_{\pm}$. Due to our 
boundary conditions, which allow only right handed solutions 
only $Q^{+}= \zeta^{\sigma} < + |-S^{56} \varepsilon_{\sigma}{}^{\tau}
\omega_{\tau}|+>$ is the solution.

{\it Gauge symmetries of action:} We must check gauge symmetries of our action, with the boundary conditions included.
We see that the Lagrange density (Eq.(\ref{actionwithb1})) {\it is not invariant under gauge transformations induced 
by $\psi'_{+} = e^{i \lambda S^{56}} \psi$}, although the boundary term itself is invariant.
The reason is that $\gamma^0 \gamma^s$, $s=5,6,$ anticommute and not commute with $S^{56}$. 
The boundary conditions allow the isometry symmetry all over the torus, except
in one only point.

{\it Conclusions:}
We start with a Weyl spinor of only one handedness in a space $M^{1+5}$,  and suggest that
the compactified space is a $S^1 \times S^1$ flat Riemann space torus with vielbeins, spin 
connection fields and torsion fields. 
We assume a ''wall'' in one point, which, under the assumption of the continuity and
differentiability of solutions of the equations of motion, makes a choice of spinors of only  
one handedness in this very point and all over the torus. 
We pay attention that  spinors manifest in $(1+3)$ ''physical'' space i) the masslessness, ii) the masslessness protected, 
iii) chirally coupling to a Kaluza-Klein gauge field, iv) through a quantized (real, nonzero) Kaluza-Klein 
gauge charge.

To be massless in $(1+3)$ space, spinors must  obey the Weyl equation on a torus: $\gamma^0\gamma^s 
f^{\sigma}{}_{s} P_{0\sigma}\psi=0, s=\{5,6\}, 
\sigma =\{(5),(6) \}$.

The requirement that the background gauge fields must obey 
the isometry   relations, makes the vielbein fields and spin connection fields on a torus constants. 
 To  have massless spinors
on a torus, the solutions of the corresponding Weyl equations must obey the periodic boundary conditions. This happens, 
if spin connection fields are proportional to integers. But since the Lagrange density for spinors, and 
consequently the corresponding Weyl equations, are not invariant under the local gauge transformations, induced by $S^{56}$,
a constant spin connection field, ''seen'' by spinors, can not be gauged away.

For any choice of spin connection fields, there exist always massless solutions of both handedness with the same
charge. This fact  spoils the massless protection mechanism. To rescue the masslessnes, we use  boundary 
conditions, which  ''distinguish'' among spinors of different handedness,  allowing spinors of only one handedness. 

Consequently massless spinors are mass protected and chirally coupled to the Kaluza-Klein gauge fields. The massless
spinors demonstrate in $(d=1+3)$ space  a charge, which is expressible with the two component
spin connection field 
$\omega_{56 \sigma}$, proportional to integers.

The ''real'' case, with  a spin in $d-$dimensional space, which should manifest in (1+3)-dimensional
''physical'' space as the spin and all the known charges,
needs, of course, more than two compactified dimensions.  All the relations are then much more complex. 
But some properties will remain, like: 
Spinors with only a spin in $1+ (d-1)$-dimensional
space will manifest in $1+ (q-1)$-dimensional space the masslessness, together with the mass protection, 
if a kind of boundary conditions
in a compactified $(d-q)-$ dimensional
space (in addition to spontaneus compactification of some dimensions, which makes then the Weyl equations in this 
compactified space to give zero on the solutions of our interest),
make possible the existence of spinors of only one handedness.

\end{document}